\documentclass[conference]{IEEEtran}

\usepackage{cite}
\usepackage{amsmath,amssymb,amsfonts}
\usepackage{algorithmic}
\usepackage{graphicx}
\usepackage{textcomp}
\usepackage{xcolor}
\usepackage{url}
\usepackage{ifthen}
\usepackage{multicol}
\usepackage{latexsym}
\usepackage{wrapfig}
\usepackage{blindtext}
\usepackage{bm}
\usepackage{amsthm}
\usepackage{ascmac}
\usepackage{lipsum}
\usepackage{mathtools}
\usepackage{cuted}
\usepackage{enumitem}
\usepackage{stfloats}
\usepackage{blindtext}
\usepackage{tikz}
\usepackage{xcolor}
\usepackage{multirow}
\newtheorem{Theorem}{Theorem}
\newtheorem{Definition}{Definition}
\newtheorem{Remark}{Remark}
\newtheorem{Lemma}{Lemma}

\ifCLASSINFOpdf
\else
\fi

\begin{document}

\title{Biometric Identification Systems With Noisy Enrollment for Gaussian Source}

%\author{\IEEEauthorblockN{Vamoua Yachongka}
%\IEEEauthorblockA{Dept. of Network and Computer Engineering\\
%The University of Electro-Communications, Tokyo\\
%Email: va.yachongka@uec.ac.jp}
%\and
%\IEEEauthorblockN{Hideki Yagi}
%\IEEEauthorblockA{Dept. of Network andComputer Engineering\\
%The University of Electro-Communications, Tokyo\\
%Email: h.yagi@uec.ac.jp}
%\and
%\IEEEauthorblockN{James Kirk\\ and Montgomery Scott}
%\IEEEauthorblockA{Starfleet Academy\\
%San Francisco, California 96678--2391\\


\author{\IEEEauthorblockN{Vamoua Yachongka\IEEEauthorrefmark{1}~~~~~~~~~~~~~~~~~
Hideki Yagi\IEEEauthorrefmark{1}~~~~~~~~~~~~~~~~~
%‪Yasutada m\IEEEauthorrefmark{1},
Yasutada Oohama\IEEEauthorrefmark{1}}
\IEEEauthorblockA{\IEEEauthorrefmark{1}Dept. of Network and Computer Engineering, The University of Electro-Communications,\\
1-5-1 Chofugaoka, Chofu, Tokyo, 182-8585 Japan.\\
Email: \{va.yachonka,h.yagi,oohama\}@uec.ac.jp}}
%Telephone: (800) 555--1212, Fax: (888) 555--1212}

\maketitle

\begin{abstract}
%In the present paper, we investigate the fundamental trade-off of identification, secrecy, storage, and privacy-leakage rates in biometric identification systems under noisy enrollment for Gaussian sources. We introduce a technique for {deriving} the capacity region of these rates by converting the system to one where the data flow is in one-way direction. Also, we provide numerical calculations of three different examples for the generated-secret model. {The numerical results imply that} it seems hard to achieve {both} high secrecy and small privacy-leakage rates {simultaneously}. In addition, as special cases, the characterization coincides with several known results in previous studies.
In the present paper, we investigate the fundamental trade-off of identification, secrecy, storage, and privacy-leakage rates in biometric identification systems for hidden or remote Gaussian sources. We introduce a technique for deriving the capacity region of these rates by converting the system to one where the data flow is in one-way direction. Also, we provide numerical calculations of three different examples for the generated-secret model. The numerical results imply that it seems hard to achieve both high secrecy and small privacy-leakage rates simultaneously. In addition, as special cases, the characterization coincides with several known results in previous studies.
\end{abstract}


% \fi
\IEEEpeerreviewmaketitle

%\textit{A full version of this paper is accessible at https://arxiv.org/abs/1902.01663}.

\section{Introduction}
The identification capacity of {biometric identification systems} (BIS) was clarified in \cite{willems} {for both discrete memoryless and Gaussian sources}.
%the authors also provided the identification capacity of the BIS for Gaussian source in \cite[Section II]{willems}.
For the discrete memoyless source (DMS), the fundamental performances of the BIS are extensively analyzed in \cite{itw},\cite{kc} for {a} visible source model (VSM) and in \cite{onur}, \cite{vy} for {a} remote source model (RSM). However, the studies under Gaussian settings are still few. For example, the optimal trade-off between secrecy and privacy-leakage was clarified in \cite{willems2009} and in order {to speed up} search complexity, hierarchical identification was taken into account in \cite{vuetal2018}. A common stand in \cite{willems2009}, \cite{vuetal2018} is that the VSM was assumed.

In this study, we extend the BIS assuming the RSM in \cite{vy} to %independent and identically distributed (i.i.d.)
Gaussian {sources}. This is motivated by the fact that the signal of biometric data (bio-data) is basically represented by vectors with continuous elements in real application and most communication links can be modeled as Gaussian channels. What is more, when the model is switched from the VSM to the {RSM}, the evaluation becomes more challenging \cite{onur}, \cite{vy} and many existing techniques for deriving the results of the VSM are not directly applicable. Thus, the extension is of both theoretical and practical {interest}. Our goal is to {find} the optimal trade-off of identification and secrecy rates in the BIS under privacy and storage constraints. We demonstrate that {an idea of converting the system to {another one} where the data flow of each user is in the same direction, {which} enables us to characterize the capacity region. More specifically, in establishing the outer bound of the region, the converted system allows us to {use} the well-known entropy power inequality} (EPI) \cite{shannon1949} twice in two opposite directions, and also its property facilitates the derivation of {the} inner bound.
%so as to recieve the capacity region.
In \cite{onur}, Mrs. {Gerber's} lemma was applied twice, too, to simplify the rate region of the RSM for binary {sources} without converting the BIS. That was possible due to the {uniformity of the source, and} the backward channel of the enrollment channel is also {the} binary symmetric channel with the same crossover probability. However, this claim is {no longer true in} {the} Gaussian case, so it is necessary {to formulate} the general behavior of the backward channel. We also provide numerical calculations of three different examples. As a consequence, we may conclude that it is difficult to achieve high secrecy and small privacy-leakage rates at the same time. To achieve {a} small privacy-leakage rate, the secrecy rate is scarified somehow. Furthermore, as {a} by-product of our result, the capacity regions of the BIS analyzed in \cite{onur} (the BIS with {a} single user) is obtained, and as special cases, it can be checked that this characterization reduces to the results given in \cite{willems}, \cite{willems2009}.

\section{System Model and Converted System}
\subsection{Notation and System {Model}}

Upper-case $A$ and lower-case $a \in \mathcal{A}$ denote random variable (RV) and its realization, {respectively}. $A^n = (A_{1},\cdots ,A_{n})$ represents a string of RVs and subscripts represent the position of a RV in the string. $f_A$ denotes the probability density function (pdf) of RV $A$. For integers $k$ and $t$ such that $k < t$, $[k:t]$ denotes the set $\{k,k+1,\cdots,t\}$.
$\log x$ stands for the natural logarithm of $x > 0$. ${\mathcal{A}^{(n)}_{\epsilon}}(\cdot)$ denotes the weakly $\epsilon$-typical set \cite{cover}, and ${{\mathcal{B}^{(n)}_{\epsilon}}}(\cdot)$ is a modified {$\epsilon$-typical} set, defined as follows.
\begin{Definition} \label{modifyset} ({Modified} {$\epsilon$-typical} set \cite[Appendix A-A]{itw})

Consider that $(X,Y,U)$ forms a Markov chain $X-Y-U$, i.e., $f_{XYU}(x,y,u)=f_{XY}(x,y)f_{U|Y}(u|y)$. The modified {$\epsilon$-typical} set ${\mathcal{B}^{(n)}_{\epsilon}}(YU)$ is defined as
\begin{align}
    &{\mathcal{B}^{(n)}_{\epsilon}}(YU) = \Big\{(y^n,u^n) : \nonumber \\
    &\Pr\{X^n\in {{\mathcal{A}^{(n)}_{\epsilon}}}(X|y^n,u^n)|(Y^n,U^n)=(y^n,u^n)\}\ge 1 - {\epsilon}\Big\},
\end{align}
where $\epsilon$ is small enough positive, and $X^n$ is drawn i.i.d. from the transition probability $\prod_{k=1}^nf_{X|Y}(x_k|y_k)$. In addition, define ${\mathcal{B}^{(n)}_{\epsilon}}(U|y^n) = \{u^n:(u^n,y^n)\in {\mathcal{B}^{(n)}_{\epsilon}}(YU)\}$ for all $y^n$, and ${\mathcal{B}^{(n)}_{\epsilon}}(U|y^n)^c$ denotes the complementary set of ${\mathcal{B}^{(n)}_{\epsilon}}(U|y^n)$.
\end{Definition}

\begin{figure}[!t]
 \centering
  \includegraphics[width = 85mm]{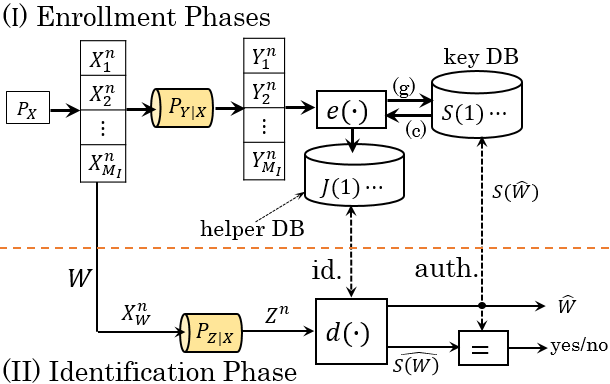}
  \vspace{-2mm}
 \caption{The generated- and chosen-secret BIS models}
 \label{fig22}
 \vspace{-5mm}
\end{figure}

The generated-secret BIS model and chosen-secret BIS model considered in this study are depicted in Fig.\ \ref{fig22}. {Arrows} (g) and (c) {indicate} the directions of the secret key of the former and {latter} models. In the former model, the secret key is extracted from the bio-data sequence, while in the {latter} one, it is chosen independently. 
%Basically, it consists of two phases: (I) {\em Enrollment Phase} and
%(I\hspace{-.1em}I) {\em Identification Phase}. We explain the details of each phase below. 
Let $\mathcal{I} = [1:M_I]$, $\mathcal{S} = [1:M_S]$, and $\mathcal{J} = [1:M_J]$ be the sets of user's indexes, secret keys, and helper data, respectively. These sets are assumed to be finite. $X^n_i$,$Y^n_i$, and $Z^n$ denote the bio-data sequence of user $i$ generated from source $P_{X}$, the output of $X^n_i$ via the enrollment channel $P_{Y|X}$, and the output of $X^n_i$ via the identification channel $P_{Z|X}$, respectively. For $i \in \mathcal{I}$ and $k \in [1:n]$, we {assume} $X_{ik} \sim \mathcal{N}(0,1)$. Note that RV with unit variance can be obtained by applying {a} scaling technique. $P_{Y|X}$ and $P_{Z|X}$ are modeled as follows:
\begin{align}
    Y_{ik} = \rho_1 X_{ik} + N_1,~~~~~Z_{k} = \rho_2 X_{ik} + N_2, \label{zxne}
\end{align}
where $|\rho_1|<1$, $|\rho_2|<1$ are the Pearson’s correlation
{coefficients}, and $N_1 \sim \mathcal{N}(0,1-\rho^2_1)$ and $N_2 \sim \mathcal{N}(0,1-\rho^2_2)$ are Gaussian RVs, independent of each other and bio-data sequences. From \eqref{zxne}, $Y$ and $Z$ are Gaussian with zero mean and unit variance, and the Markov chain $Y-X-Z$ holds. Then, the pdf corresponding to the {tuple} $(X^n_i,Y^n_i,Z^n)$ is given by
\begin{align}
    f_{X^n_iY^n_iZ^n}(x^n_i,y^n_i,z^n)
    = \textstyle\prod_{\substack{k=1}}^nf_{XYZ}(x_{ik},y_{ik},z_{k}),
    %= \prod_{\substack{k=1}}^nf_{Y|X}(y_{k}|x_{k})\cdot f_{X}(x)\cdot f_{Z|X}(z_{k}|x_{k}),
\end{align}
where for $x,y,z \in \bm{R}$,
\begin{align}
    \hspace{-2mm} f_{XYZ}(x,y,z) &= f_{X}(x)\cdot f_{Y|X}(y|x)\cdot f_{Z|X}(z|x), \label{expan1} \\
    &=  \frac{1}{\sqrt{(2\pi)^3(1-\rho^2_1)(1-\rho^2_2)}}\nonumber \\
    &\hspace{-12mm}~~~\cdot \exp{\left(-\left(\frac{x^2}{2} + \frac{(y-\rho_1x)^2}{2(1-\rho^2_1)}+\frac{(z-\rho_2x)^2}{2(1-\rho^2_2)}\right)\right)}. \label{jdisgs}
\end{align}

%each symbol of $(X^n_i,Y^n_i,Z^n)$ is i.i.d. according to    the identification channel $Z^n$ (bio-data sequence of individual $i$) with realizations on $\mathcal{X}^n$ is generated i.i.d. from a source $P_X$.  and  are memoryless channels at the encoder and decoder, observing $Y^n_i$ with realizations on $\mathcal{Y}^n$ and $Z^n$ with realizations on $\mathcal{Z}^n$, respectively, when $X^n_i$ is input. The joint distribution among $X^n_i$, $Y^n_i$, and $Z^n$ are given by

In the generated-secret BIS model, upon observing $Y^n_i$, the encoder $e$ generates secret key $S(i) \in \mathcal{S}$ and helper data $J(i) \in \mathcal{J}$ as $(S(i),J(i)) = e(Y^n_i)$. $J(i)$ is stored at position $i$ in public database (helper DB) and $S(i)$ is {saved} in key DB, {which is installed in a secure location}. Seeing $Z^n$, the decoder $d$ estimates $(\widehat{W},\widehat{S(W)})$ from $Z^n$ and all helper data in DB $\bm{J} \equiv \{J(1),\cdots,J(M_I)\}$, i.e., $(\widehat{W},\widehat{S(W)})=d(Z^n,\bm{J})$.
In the chosen-secret BIS model, $S(i)$ is chosen uniformly from $\mathcal{S}$ and independent of other RVs. The encoder forms the helper data by $J(i) = e(Y^n_i,S(i))$ for every individual. The decoder $d$ owns the same functionality as the {generated-secret} model.

\subsection{Converted System}
\begin{figure}[!t]
 \centering
  \includegraphics[width = 85mm]{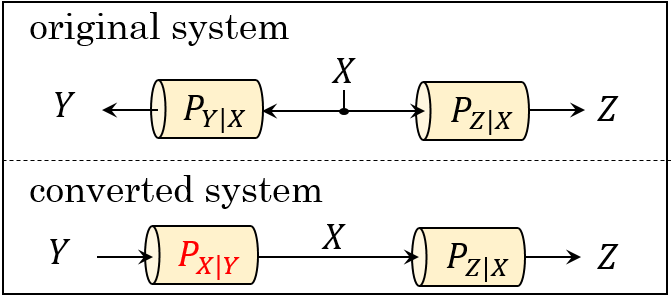}
  \vspace{-2mm}
 \caption{The original (top) and converted (bottom) systems}
 \label{fig33}
 \vspace{-5mm}
\end{figure}

The original system, having $X$ as input source and $Y,Z$ as outputs, is {in} the top figure in Fig.\ \ref{fig33}. There are two main obstacles toward characterizing the capacity regions directly from this system. (I) In establishing the converse proof, {an upper} bound regarding RV $Y$ for a fixed condition of RV $X$ is needed, but it is {laborious} to pursue the desired bound since applying EPI to the first relation in \eqref{zxne} only produces a lower {bound}. (II) It seems difficult to prove the achievability part based on generating auxiliary sequences from edge $X$, e.g., the rate settings. To overcome these bottlenecks, we introduce an idea of converting the original system to a new one in which the data flow of each user is one-way from $Y$ to $Z$ without losing its general properties.
%to a one, which the data flows of each user are in the same direction.
The image of this idea is shown in the bottom figure of Fig.\ \ref{fig33}, where $Y$ becomes input virtually. To achieve this objective, knowing the property of the backward channel $P_{X|Y}$, namely, how $X$ correlates to the virtual input $Y$, is crucial and we explore that in the rest of this section.
%Here, we introduce an idea of converting the original system to a new one where the data flow of each user is in one-way direction from $Y$ to $Z$ without losing its general properties. to a one, which the data flows of each user are in the same direction. The image of this idea is shown in the bottom figure of Fig.\ \ref{fig33}, where $Y$ becomes input virtually. This concept enables us to overcome the bottlenecks toward characterizing the capacity region. toward is necessary for characterizing the capacity region of the BIS for Gaussian source. More specifically, in establishing the converse proof, we can derive a tight upper bound regarding RV $Y$ for a fixed condition of RV $X$. It is impossible to pursue that bound from the original system since applying EPI to the first relation in \eqref{zxne} will only receives a lower bound. However, the converted system provides  the desirable Also, it allows us to consider generating auxiliary sequences from edge $Y$ in the achievability. To overcome this bottleneck, one has to transform the original system into the one like bottom figure of Fig.\ \ref{fig33}. To achieve this objective, knowing how $X$ correlates to the virtual input $Y$ is crucial and we explore it in the rest of this section.

Due to the Markov chian $Y-X-Z$, \eqref{expan1} can also be expanded in the following form.
\begin{align}
    f_{XYZ}(x,y,z) = f_{Y}(y)\cdot f_{X|Y}(x|y)\cdot f_{Z|X}(z|x). \label{convertedsystem}
\end{align}
Observe that
\begin{align}
    \frac{x^2}{2} + \frac{(y-\rho_1x)^2}{2(1-\rho^2_1)} &= \frac{x^2}{2} + \frac{y^2}{2(1-\rho^2_1)} - \frac{\rho_1xy}{1-\rho^2_1} +  \frac{(\rho_1x)^2}{2(1-\rho^2_1)} \nonumber \\
    &= \frac{y^2}{2} + \frac{(x-\rho_1y)^2}{2(1-\rho^2_1)}.
\end{align}

Without loss of generality, the exponential part in \eqref{jdisgs} can be rearranged as
\begin{align}
    -\left(\frac{y^2}{2} + \frac{(x-\rho_1y)^2}{2(1-\rho^2_1)}+\frac{(z-\rho_2x)^2}{2(1-\rho^2_2)}\right). \label{jdisgs2}
\end{align}
From \eqref{convertedsystem} and \eqref{jdisgs2}, we may conclude that the following {equations} hold.
\begin{align}
    X_{ik} &= \rho_1 Y_{ik} + N'_1, \label{xyne}\\
    Z_{k} &= \rho_2 X_{ik} + N_2 = {\rho_1\rho_2}Y_{ik} + \rho_2N'_1 + N_2 \label{zxni2}
\end{align}
with some RV $N'_1 \sim \mathcal{N}(0,1-\rho^2_1)$. Equation \eqref{xyne} describes the output of the backward channel with having $Y$ as input. The above relations play key roles for solving the problem of the RSM, and indeed we use them in many steps {during the analysis} in this study. In \cite{willems2009} and \cite{vuetal2018}, the concept of this transformation is not seen because the {enrollment} channel does not exist due to the assumption of VSM as mentioned before.
\begin{Remark} \label{xyzxy}
In case there is no operation of scaling, equations \eqref{xyne} and \eqref{zxni2} are settled as follows. Suppose that $X_{ik}\sim \mathcal{N}(0,\sigma^2_x)$, $Y_{ik} = X_{ik} + D_1$, and $Z_{k} = X_{ik} + D_2$, where $D_1 \sim \mathcal{N}(0,\sigma^2_1)$ and $D_2 \sim \mathcal{N}(0,\sigma^2_2)$ are Gaussian RVs, and independent of other RVs. By applying the arguments around \eqref{convertedsystem}--\eqref{jdisgs2}, we obtain that
\begin{align}
    X_{ik} &= \frac{\sigma^2_x}{\sigma^2_x + \sigma^2_1} Y_{ik} + D'_1 \label{xy}  \\
    Z_{k} &= X_{ik} + N'_2 = \frac{\sigma^2_x}{\sigma^2_x + \sigma^2_1} Y_{ik} + D'_1 + D_2, \label{zx}
\end{align}
where $D'_1 \sim \mathcal{N}(0,\frac{\sigma^2_x \sigma^2_1}{\sigma^2_x + \sigma^2_1})$ is Gaussian and independent of other RVs. The capacity regions of the models {considered} in this study can also be characterized from \eqref{xy} and \eqref{zx}. However, equation developments need more space and it does not look so neat. Herein, we pursue our results based on the method that RVs $X$, $Y$, and $Z$ are standardized.
%(cf.\ \eqref{xyne} and \eqref{zxni2}).
\end{Remark}

Now from \eqref{xyne} and \eqref{zxni2}, it is not difficult to calculate that
\begin{align}
    I(X;Y) &= \frac{1}{2}\log\left(\frac{1}{1-\rho^2_1}\right), \label{ixy} \\
    I(Z;Y) &= \frac{1}{2}\log\left(\frac{1}{1-\rho^2_1\rho^2_2}\right), \label{izy}
\end{align}
{where \eqref{izy} is attained because the variance of the noise term $\rho_2N'_1 + N_2$ in \eqref{zxni2} is equal to $1-\rho^2_1\rho^2_2$}.
\section{Statement of Results}
In this section, we provide the formal definitions of both the generated- and chosen-secret BIS models, and state the main results.
%After that, we take a look into how these results links to the ones characterized in previous studies.
\subsection{Problem Formulation and Main Results}
The achievability definition for the generated-secret BIS model is given below.
\begin{Definition} \label{def11}

A tuple of identification, secrecy, storage, and privacy-leakage rates $(R_I,R_S,R_J,R_L)$ is said to be achievable for {a} Gaussian source if for any $\delta > 0$ and large enough $n$ there exist pairs of encoders and decoders satisfying
\begin{align}
  \textstyle \max_{i \in \mathcal{I}}\textstyle\Pr\{(\widehat{W},\widehat{S(W))} \neq (W,S(W)) &|W=i\} \leq  \delta, \label{errorp} \\
  \textstyle \frac{1}{n}\log M_I &\geq R_I - \delta{,} \label{id} \\
  %\textstyle \frac{1}{n}\log M_C &\geq R_C - \delta{,} \label{c} \\
  \textstyle \min_{i \in \mathcal{I}}\frac{1}{n}H(S(i)) &\geq R_S - \delta, \label{secretk} \\
  \textstyle \frac{1}{n}\log{M_J} &\leq R_J + \delta, \label{storage} \\
    \textstyle \max_{i \in \mathcal{I}}\frac{1}{n}I(S(i);J(i)) &\leq \delta. \label{secrecy} \\
  \textstyle \max_{i \in \mathcal{I}}\frac{1}{n}I(X^n_i;J(i)) &\leq R_L + \delta. \label{privacy}
  %\textstyle \max_{i \in \mathcal{I}}\frac{1}{n}I(S_C(i);S_G(i)) &\leq {\delta}, \label{g} \\
\end{align}
Moreover, $\mathcal{R}_G$ is defined as the set of all achievable rate tuples for the generated-secret BIS model, called the capacity region.
\qed
\end{Definition}

The achievability definition for the chosen-secret BIS model is provided as follows:
%({Achievability} for the chosen-secret BIS model)
\begin{Definition} \label{def22}
A tuple $(R_I,R_S,R_J,R_L)$ is said to be achievable for {a} Gaussian source if there exist pairs of encoders and decoders that satisfy all the requirements in Definition \ref{def11} for any $\delta > 0$ and large enough $n$. In addition, $\mathcal{R}_C$ is defined as the capacity region of {the chosen-secret BIS} model.
\qed
\end{Definition}
Note that the {left-hand} side of \eqref{secretk} is {expressed as} $\frac{1}{n}\log M_S$ because $S(i)$ is chosen uniformly from $\mathcal{S}$.

\begin{Remark}
It is worthwhile to mention that it is not really suitable to call the {rate} of helper data the storage rate. In \cite{vy}, it was called the template rate instead and the reason behind the scene is that there exist two databases in the BIS, namely, databases of secret keys and helper data or templates. The storage space of the database for storing the templates is minimized, while that {for} the secret keys is maximized. Thus, only a part of the entire storage space of the BIS is being minimized. In this paper, however, we also use {this} term because it is widely used in many previous works, {e.g., \cite{kc}, \cite{onur}}.
%so as to make the readers feel more comfortable to follow the contents.
\end{Remark}

Now we are ready to present our main results.
\begin{Theorem} \label{th1}
The capacity regions for the generated- and chosen-secret BIS {models} are given by
\begin{align}
\mathcal{R}_G = \{(R_I,&R_S,R_J,R_L): \nonumber \\
R_I + R_S &\le \frac{1}{2}\log\left(\frac{1}{\alpha\rho^2_1\rho^2_2 + 1 - \rho^2_1\rho^2_2}\right), \nonumber \\
R_J &\geq \frac{1}{2}\log\left(\frac{\alpha\rho^2_1\rho^2_2 + 1 - \rho^2_1\rho^2_2}{\alpha}\right) + R_I,\nonumber \\
R_L &\ge \frac{1}{2}\log\left(\frac{\alpha\rho^2_1\rho^2_2 + 1 - \rho^2_1\rho^2_2}{\alpha\rho^2_1 + 1 - \rho^2_1}\right) + R_I, \nonumber \\
R_I &\geq 0, R_S\geq 0~{\rm for}~{\rm some}~0<\alpha\le 1\}, \label{theorem1} \\
\mathcal{R}_C = \{(R_I,&R_S,R_J,R_L): \nonumber\\
R_I + R_S &\le \frac{1}{2}\log\left(\frac{1}{\alpha\rho^2_1\rho^2_2 + 1 - \rho^2_1\rho^2_2}\right), \nonumber \\
R_J &\ge \frac{1}{2}\log\left(\frac{1}{\alpha}\right), \nonumber \\
R_L &\ge \frac{1}{2}\log\left(\frac{\alpha\rho^2_1\rho^2_2 + 1 - \rho^2_1\rho^2_2}{\alpha\rho^2_1 + 1 - \rho^2_1}\right) + R_I, \nonumber \\
R_I &\geq 0, R_S\geq 0~{\rm for}~{\rm some}~0<\alpha\le 1\}. \label{theorem2}
\end{align}
\qed
\end{Theorem}
Similar to a conclusion in \cite{vy}, the {lower} bound {on} $R_J$ in $\mathcal{R}_C$ is greater than the one in $\mathcal{R}_G$. This means the chosen-secret BIS model consumes more storage space. This is because the information related to the secret key chosen at the encoder must be saved together with the helper data in DB {so as to aid the estimation of the key} at the decoder. Unlike $R_J$, the bound {on} $R_L$ remains unchanged in both models, and it rises in accordance with the increase of $R_I$.

As a by-product of Theorem \ref{th1}, the following {remark} is obtained.
\begin{Remark} \label{coro1} 
The capacity regions of the generated- and chosen-secret BIS models with {a} single user (the models considered in \cite{onur}) for Gaussian {sources} are given by substituting $R_I=0$ into $\mathcal{R}_G$ and $\mathcal{R}_C$, respectively.
\end{Remark}
The proofs of Remark \ref{coro1} can be done similarly to the arguments {in} proving Theorem \ref{th1}.

As special cases, when $R_S=0$, and $R_J$ and $R_L$ are large enough ($R_J,R_L\rightarrow \infty$), the maximum value of $R_I$ is $\frac{1}{2}\log(\frac{1}{1 - \rho^2_1\rho^2_2})$. This value is exactly the identification capacity $I(Y;Z)$ (cf.\ \eqref{izy}) derived in \cite{willems}, and it is achieved when $\alpha \downarrow 0$. Moreover, when $R_I=0$, $R_J\rightarrow \infty$, and the enrollment channel is noiseless $(\rho_1 = 1)$, one can see that Theorem \ref{th1} naturally reduces to the characterizations of \cite{willems2009}. 

\begin{Remark}
{If there is no scaling as in} \eqref{xy} and \eqref{zx} in Remark \ref{xyzxy}, {the capacity regions of the generated- and chosen-secret BIS models $\mathcal{R}'_G $ and $\mathcal{R}'_C$, respectively, are characterized as follows}:
\begin{align}
\mathcal{R}'_G = \{(R_I,&R_S,R_J,R_L): \nonumber \\
R_I + R_S &\le \frac{1}{2}\log\left(\frac{(\sigma^2_x + \sigma^2_1)(\sigma^2_x+\sigma^2_2)}{\alpha\sigma^4_x + \sigma^2_x\sigma^2_1 + \sigma^2_1\sigma^2_2 + \sigma^2_2\sigma^2_x}\right), \nonumber \\
R_J &\geq \frac{1}{2}\log\left(\frac{\alpha\sigma^4_x + \sigma^2_x\sigma^2_1 + \sigma^2_1\sigma^2_i + \sigma^2_2\sigma^2_x}{\alpha(\sigma^2_x + \sigma^2_1)(\sigma^2_x+\sigma^2_2)}\right) + R_I,\nonumber \\
R_L &\ge \frac{1}{2}\log\left(\frac{\alpha\sigma^4_x + \sigma^2_x\sigma^2_1 + \sigma^2_1\sigma^2_2 + \sigma^2_2\sigma^2_x}{(\alpha\sigma^2_x + \sigma^2_1)(\sigma^2_x+\sigma^2_2)}\right) + R_I, \nonumber \\
R_I &\geq 0, R_S\geq 0~{\rm for}~{\rm some}~0<\alpha\le 1\}, \label{theorem11} \\
\mathcal{R}'_C = \{(R_I,&R_S,R_J,R_L): \nonumber\\
R_I + R_S &\le \frac{1}{2}\log\left(\frac{(\sigma^2_x + \sigma^2_1)(\sigma^2_x+\sigma^2_2)}{\alpha\sigma^4_x + \sigma^2_x\sigma^2_1 + \sigma^2_1\sigma^2_2 + \sigma^2_2\sigma^2_x}\right), \nonumber \\
R_J &\ge \frac{1}{2}\log\left(\frac{1}{\alpha}\right), \nonumber \\
R_L &\ge \frac{1}{2}\log\left(\frac{\alpha\sigma^4_x + \sigma^2_x\sigma^2_1 + \sigma^2_1\sigma^2_2 + \sigma^2_2\sigma^2_x}{(\alpha\sigma^2_x + \sigma^2_1)(\sigma^2_x+\sigma^2_2)}\right) + R_I, \nonumber \\
R_I &\geq 0, R_S\geq 0~{\rm for}~{\rm some}~0<\alpha\le 1\}. \label{theorem22}
\end{align}
It can be verified that $\mathcal{R}_G$ and $\mathcal{R}_C$ are equivalent to  $\mathcal{R}'_G$ and $\mathcal{R}'_C$, respectively, if we set $\rho^2_1 = \frac{\sigma^2_x}{\sigma^2_x + \sigma^2_1}$ and $\rho^2_2 = \frac{\sigma^2_x}{\sigma^2_x + \sigma^2_2}$.
\end{Remark}

\subsection{Examples}
For the sake of succinct discussion, we only concentrate on the generated-secret BIS model at which $R_I = 0$.
%The difference between $\mathcal{R}_G$ and $\mathcal{R}_C$ is the bound of $R_J$. If we explore the , for example, the bound of $R_L$ and the bound of the sum of $R_I$ and $R_S$, exploring the behaviors of $\mathcal{R}_G$ alone supply enough information to describe those of $\mathcal{R}_C$, too.
We first look over some special points of secrecy and privacy-leakage rates when storage rate becomes extremely low or large. We first define two rate functions
\begin{align}
    R^*_S(R_J) &=\max_{(R_S,R_J,R_L)\in\mathcal{R}_G}R_S, \label{rsalpha} \\
    R^*_L(R_J) &=\min_{(R_S,R_J,R_L)\in\mathcal{R}_G}R_L, \label{rlalpha}
\end{align}
where \eqref{rsalpha} and \eqref{rlalpha} are the maximum secrecy rate and minimum privacy-leakage rate, respectively, for given $R_J$. Moreover, we define 
$
    R^{\alpha}_J = \frac{1}{2}\log(\frac{\alpha\rho^2_1\rho^2_2 + 1 - \rho^2_1\rho^2_2}{\alpha})
$
so that we can write
\begin{align}
    R^*_S(R^{\alpha}_J) &= \frac{1}{2}\log\left(\frac{1-\rho^2_1\rho^2_2/2^{2(R^{\alpha}_J)}}{1 - \rho^2_1\rho^2_2}\right), \label{rsrho1rho2rj}
\end{align}
\begin{align}
    R^*_L(R^{\alpha}_J) &= \frac{1}{2}\log\left(\frac{1 - \rho^2_1\rho^2_2}{1- \rho^2_1 + \rho^2_1(1-\rho^2_2)/2^{2(R^{\alpha}_J)}}\right). \label{rlrho1rho2rj}
\end{align}

As $R^{\alpha}_J \rightarrow \infty~(\alpha \downarrow 0)$, the optimal asymptotic secrecy rate and the quantity of privacy-leakage approach to
\begin{align}
    \lim_{R^{\alpha}_J \rightarrow \infty}R^*_S(R^{\alpha}_J) &=\frac{1}{2}\log\left(\frac{1}{1 - \rho^2_1\rho^2_2}\right) = I(Y;Z), \label{limrs} \\
    \lim_{R^{\alpha}_J \rightarrow \infty}R^*_L(R^{\alpha}_J) &=\frac{1}{2}\log\left(\frac{1 - \rho^2_1\rho^2_2}{1- \rho^2_1}\right)\nonumber \\
    &=\frac{1}{2}\log\left(\frac{1}{1- \rho^2_1}\right) -  \frac{1}{2}\log\left(\frac{1}{1 - \rho^2_1\rho^2_2}\right)\nonumber\\
    &= I(X;Y)-I(Z;Y). \label{limrl}
\end{align}
The result \eqref{limrs} corresponds to the optimal asymptotic secrecy rate \cite[Sect. III-B]{willems2009} and in order to achieve this rate, it is required to take the storage rate to infinity and leak the user's privacy {up to} rate $I(X;Y)-I(Z;Y)$.

In contrast, when $R_J \downarrow 0$, it is evident that $R_S$ and $R_L$ become zero as well, which does not carry much information. However, {to investigate} the BIS that achieves high secrecy and small privacy-leakage rates in {the} low storage rate regime, the zero-rate slopes of secrecy and privacy-leakage rates, namely, how fast they converge to zero, are important indicators. In views of \eqref{rsrho1rho2rj} and \eqref{rlrho1rho2rj}, by a few steps of calculations, the slopes of secrecy and privacy-leakage rates at $R_J \downarrow 0$ can be determined as follows:
\begin{align}
    \frac{dR^{*}_S(R^{\alpha}_J)}{dR^{\alpha}_J}\bigg|_{R^{\alpha}_J=0} &= \frac{\rho^2_1\rho^2_2}{1-\rho^2_1\rho^2_2}, \label{slopers} \\
    \frac{dR^*_L(R^{\alpha}_J)}{dR^{\alpha}_J}\bigg|_{R^{\alpha}_J=0} &= \frac{\rho^2_1(1-\rho^2_2)}{1-\rho^2_1\rho^2_2} = \frac{\rho^2_1\rho^2_2}{1-\rho^2_1\rho^2_2}\cdot\frac{1-\rho^2_2}{\rho^2_2}, \label{sloperl}
\end{align}
where \eqref{slopers} is equal to the signal-to-noise ratio of the compound channel from $Y$ to $Z$. This value multiplied by the reverse of the signal-to-noise ratio of {the channel $P_{Z|X}$} appears in the slope of privacy-leakage rate in \eqref{sloperl}.

\begin{figure*}[b!]
\centering
\begin{minipage}{.5\textwidth}
  \centering
  \includegraphics[width=.95\linewidth]{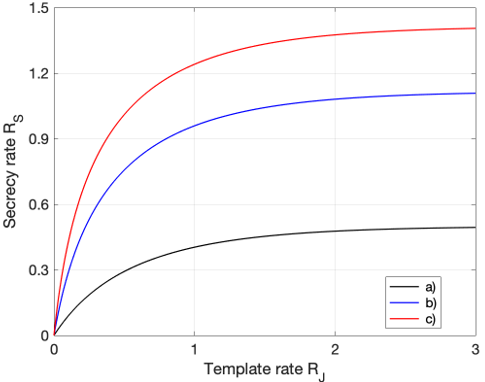}
  \caption{The projection of the rate region onto $R_JR_S$-plane for Ex.\ 2.}
  \label{rjrs-g1}
 \vspace{3mm}
\end{minipage}%
\begin{minipage}{.5\textwidth}
  \centering
  \includegraphics[width=.95\linewidth]{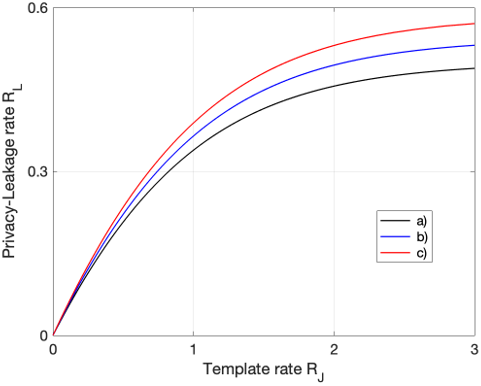}
  \caption{The projection of the rate region onto $R_JR_L$-plane for Ex.\ 2.}
  \label{rjrl-g1}
  \vspace{3mm}
\end{minipage}
\begin{minipage}{.5\textwidth}
  \centering
  \includegraphics[width=.95\linewidth]{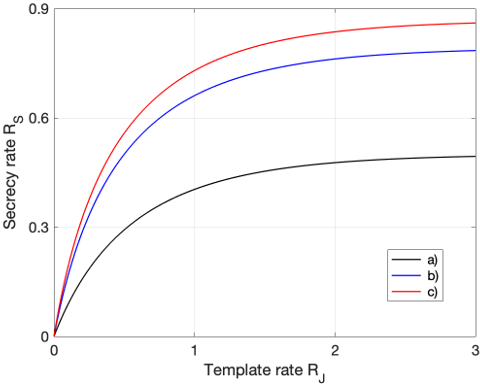}
  \caption{The projection of the rate region onto $R_JR_S$-plane for Ex.\ 3.}
  \label{rjrs-g2}
\end{minipage}%
\begin{minipage}{.5\textwidth}
  \centering
  \includegraphics[width=.95\linewidth]{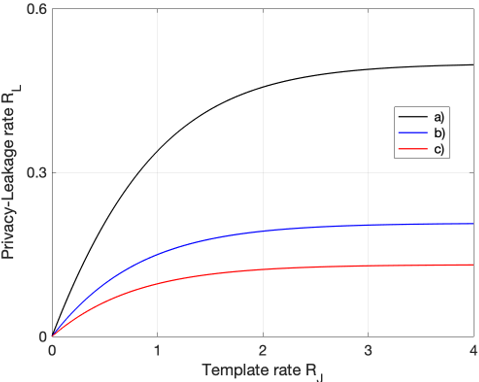}
  \caption{The projection of the rate region into $R_JR_L$-plane for Ex.\ 3.}
  \label{rjrl-g2}
\end{minipage}
\end{figure*}

Next, we give numerical computations of three different examples and take a look into behaviors of the special points.
\begin{enumerate}[label= Ex.\ \arabic*:]
\setlength{\itemindent}{3mm}
    \item a) $\rho^2_1=3/4,\rho^2_2=2/3$,~b) $\rho^2_1=7/8,\rho^2_2=2/3$, \\
    c) $\rho^2_1=15/16,\rho^2_2=2/3$,
    \item a) $\rho^2_1=3/4,\rho^2_2=2/3$,~b) $\rho^2_1=9/10,\rho^2_2=7/8$,\\
    c) $\rho^2_1=15/16,\rho^2_2=11/12$,
    \item a) $\rho^2_1=3/4,\rho^2_2=2/3$,~b) $\rho^2_1=3/4,\rho^2_2=8/9$,\\
    c) $\rho^2_1=3/4,\rho^2_2=14/15$.
\end{enumerate}
Note that as $\rho^2_1,\rho^2_2$ are large, the noises {added to the bio-data sequences} at encoder and decoder become small. Example 1 is the case where the noise at encoder is gradually small from a) to c), but the noise at the decoder stays constant for each round. Example 2 is the case in which the noises at both encoder and decoder are improved gradually from a) to c). Example 3 is opposite to Example 1. The calculated results of the {secrecy} and privacy-leakage rates for these cases are summarized in Table \ref{table1} and \ref{table2}, {and Fig.\ \ref{rjrs-g1}--\ref{rjrl-g2}}.

\begin{table}[!ht]
\caption{The secrecy and privacy-leakage rates when $R_J \rightarrow \infty$.}
\begin{center}
\def\arraystretch{1.2}
\begin{tabular}{ |c|c|c|c|c|c|c| } 
%\hline
%\multicolumn{5}{|c|}{Example 1}\\
\hline
\multirow{2}{*}{Cases} & \multicolumn{3}{c|}{Secrecy rate} & \multicolumn{3}{c|}{Privacy-Leakage rate}\\
\cline{2-7} & a) & b) & c) & a) & b) & c) \\
\hline
Ex.\ 1 & $0.5$ & $0.63$ & $0.70$  & $0.5$& $0.87$& $1.29$\\ 
\hline
Ex.\ 2 & $0.5$ & $1.12$ & $1.41$  & $0.5$ & $0.54$ & $0.59$\\ 
\hline
Ex.\ 3 & $0.5$ & $0.79$ & $0.87$  & $0.5$ & 0.20 & 0.13\\ 
\hline
\end{tabular}
\label{table1}
\vspace{2mm}
\caption{The slopes of secrecy and privacy-leakage rates at $R_J \downarrow 0$.}
\begin{tabular}{ |c|c|c|c|c|c|c| } 
%\hline
%\multicolumn{5}{|c|}{Example 1}\\
\hline
\multirow{2}{*}{Cases} & \multicolumn{3}{c|}{Secrecy rate} & \multicolumn{3}{c|}{Privacy-Leakage rate}\\
\cline{2-7} & a) & b) & c) & a) & b) & c) \\
\hline
Ex.\ 1 & $1.0$ & $1.40$ & $1.67$  & $0.5$& $0.7$& $0.83$\\ 
\hline
Ex.\ 2 & $1.0$ & $3.71$ & $6.11$  & $0.5$ & $0.53$ & $0.56$\\ 
\hline
Ex.\ 3 & $1.0$ & $2.0$ & $2.33$  & $0.5$ & $0.25$ & $0.17$\\ 
\hline
\end{tabular}
\label{table2}
\end{center}
\vspace{-4mm}
\end{table}
%\vspace{-5mm}

It is ideal to keep the privacy-leakage rate small, while produce high secrecy rate, but Example 1 works out in the opposite way {(cf.\ the rows of Ex. 1 in Table I and II)}, so this is not a preferable choice.  Example 2 {realizes a} high secrecy rate, but the amount of privacy-leakage remains high at some level, too {(cf.\ the rows of Ex.\ 2 in Table I and II, and Fig. \ref{rjrs-g1} and \ref{rjrl-g1}}). On the other hand, in Example 3, {the} privacy-leakage rate declines, but the secrecy rate becomes small compared to Example 3 {(cf.\ the rows of Ex.\ 3 in Table I and II, and Fig. \ref{rjrs-g2} and \ref{rjrl-g2}}). From these behaviors, we may conclude that it is {unmanageable} to achieve {both} a high secrecy rate and small privacy-leakage at the same time. If one aims to achieve {a} high secrecy rate, it is important to diminish the noises at both encoder and decoder, e.g., deploying quantizers with high quality, but this could result in leaking more user's privacy. In different circumstances, to achieve {a} small privacy-leakage rate, it is preferable to maintain a certain level of noise at encoder and pay sufficient {attention} for processing the noise at decoder. In this way, however, the gain of the secrecy rate may be dropped.

\section{Proof of the Region $\mathcal{R}_G$} \label{pft}
In this section, we give the proof of the capacity region {of the} generated-secret BIS model.
%We omit the derivation of $\mathcal{R}_C$ since it follows by applying the one-time pad operation to conceal the key. For those who are interested, the same technique is used in \cite{itw},\cite{onur},\cite{vy}.
%The proof begins with the converse part and follows by the achievability.
\subsection{Converse Part}
We consider a more relaxed case where $W$ {is uniformly distributed on $\mathcal{I}$}, and (\ref{errorp}) is replaced with the average error criterion
$
\Pr\{(\widehat{W},\widehat{S(W)})\neq (W,S(W))\} \leq \delta.
$
We shall show that the capacity region for this case, which contains $\mathcal{R}_G$, is contained in (\ref{theorem1}). We assume that a rate tuple $(R_I,R_S,R_J,R_L)$ is achievable.

\medskip
\noindent{{\em Analysis of Secrecy Rate}}:
We begin with considering the joint entropy of $W$ and $S(W)$ as
\begin{align}
H&(W,S(W))\nonumber \\
&= H(W,S(W)|Z^n,\bm{J}) + I(W,S(W);Z^n,\bm{J}) \nonumber \\
&\overset{\mathrm{(a)}}{=} H(W,S(W)|\widehat{W},\widehat{S(W)},Z^n,\bm{J}) + I(W,S(W);\bm{J}) \nonumber \\
&~~~~+ I(W,S(W);Z^n|\bm{J}) \nonumber \\
&\overset{\mathrm{(b)}}{\le} H(W,S(W)|\widehat{W},\widehat{S(W)}) + I(W,S(W);J(W)) \nonumber \\
&~~~~+ I(W,S(W);Z^n|J(W)) \nonumber \\
&\overset{\mathrm{(c)}}{\le} n\delta_n + I(W;J(W)) + I(S(W);J(W)|W) \nonumber \\
&~~~~+ I(W,S(W);Z^n|J(W)) \nonumber \\
&\overset{\mathrm{(d)}}{\le} n(\delta_n + \delta) + h(Z^n|J(W))- h(Z^n|J(W),S(W)) \nonumber \\
&\overset{\mathrm{(e)}}{\le} n(\delta_n + \delta) + h(Z^n)- h(Z^n|J(W),S(W)), \label{hzjwsw}
\end{align}
where
\begin{enumerate}[label=(\alph*)]
	\setcounter{enumi}{0}
	\item holds since $(\widehat{W},\widehat{S(W)})$ {is a} function of $(Z^n,\bm{J})$,
	\item follows because conditioning reduces entropy, and only $J(W)$ is possibly dependent on $Z^n$ and $S(W)$,
    \item is due to Fano's inequality {with $\delta_n = \frac{1}{n}(1 +\delta\log M_IM_S)$},
	\item follows since $W$ is independent of other RVs and \eqref{secrecy} is applied,
    \item follows because conditioning reduces entropy.
\end{enumerate}
Also, since $W$ is uniformly distributed on $\mathcal{I}$, we have that
\begin{align}
    H(W,S(W)) &= H(W) + H(S(W)|W) \nonumber\\
    &\ge \log M_I + \min_{w \in \mathcal{I}}H(S(w)). \label{hwsw}
\end{align}
From \eqref{id}, \eqref{secretk}, \eqref{hzjwsw}, and \eqref{hwsw}, it yields that
\begin{align}
    R_I + R_S \le h(Z) - \frac{1}{n}h(Z^n|J(W),S(W)) + 3\delta + \delta_n. \label{rs2}
\end{align}

\medskip
\noindent{\em {Analysis of Storage Rate}}:
\vspace{-2mm}
\begin{align}
    n(R_J + \delta) &\ge \log M_J \ge \max_{w \in \mathcal{I}}H(J(w)) \ge H(J(W)|W) \nonumber \\
    &= I(Y^n_W;J(W)|W) = I(Y^n_W,W;J(W)) \nonumber \\
    &= I(Y^n_W,W,S(W);J(W)) \nonumber \\
    &~~~~-I(S(W);J(W)|Y^n_W,W) \nonumber \\
    &\overset{\mathrm{(f)}}{=} h(Y^n_W,W,S(W))-h(Y^n_W,W,S(W)|J(W)) \nonumber \\
    &\overset{\mathrm{(g)}}{=} h(Y^n_W)+H(W)-H(W,S(W)|J(W)) \nonumber \\
    &~~~~- h(Y^n_W|J(W),S(W)) \nonumber \\
    &\overset{\mathrm{(h)}}{\ge} h(Y^n_W) + \log M_I -H(W,S(W)) \nonumber \\
    &~~~~- h(Y^n_W|J(W),S(W)) \nonumber \\
    &\overset{\mathrm{(i)}}{\ge} h(Z^n|J(W),S(W))- h(Y^n_W|J(W),S(W))  \nonumber \\
    &~~~+ nR_I -n(\delta_n + 2\delta), \label{rj2}
\end{align}
where
\begin{enumerate}[label=(\alph*)]
	\setcounter{enumi}{5}
        \item hold as $(S(W),J(W))$ {is} a function of $Y^n_W$,
	    \item follows since and $W$ is independent of other RVs, and $S(W)$ is a function of $Y^n_W$,
        \item follows because conditioning reduces entropy and $W$ is uniformly distributed on $\mathcal{I}$,
        \item follows because $h(Y^n_W)=h(Z^n)={\frac{n}{2}}\log(2 \pi e)$, and \eqref{id} and \eqref{hzjwsw} are applied.
\end{enumerate}

\medskip
\noindent{\em {Analysis of Privacy-Leakage Rate}}:
\begin{align}
n(&R_L + \delta) \nonumber \\
&\ge \max_{w \in \mathcal{I}}I(X^n_w;J(w)) \ge I(X^n_W;J(W)|W) \nonumber \\
&= I(X^n_W,W;J(W)) \nonumber \\
&= I(X^n_W,W,S(W);J(W)) - I(S(W);J(W)|X^n_W,W) \nonumber \\
&\ge h(X^n_W,W,S(W)) - h(X^n_W,W,S(W)|J(W)) \nonumber \\
&~~~~- H(S(W)|X^n_W) \nonumber \\
&\ge h(X^n_W) + H(W) + H(S(W)|X^n_W)  \nonumber \\
&~~~~- H(W,S(W)|J(W))- h(X^n_W,|J(W),S(W)) \nonumber \\
&~~~~ - H(S(W)|X^n_W) \nonumber \\
&\overset{\mathrm{(j)}}{\ge} h(X^n_W) + \log M_I - H(W,S(W)) \nonumber \\
&~~~~- h(X^n_W,|J(W),S(W)) \nonumber \\
&\overset{\mathrm{(k)}}{\ge} h(Z^n|J(W),S(W))- h(X^n_W|J(W),S(W)) \nonumber \\
&~~~~+ nR_I -n(\delta_n + 2\delta),
\label{rl2}
\end{align}
where
\begin{enumerate}[label=(\alph*)]
	\setcounter{enumi}{9}
        \item follows as conditioning reduces entropy and $W$ is uniformly distributed on $\mathcal{I}$,
        \item follows because $h(X^n_W)=h(Z^n)$, and \eqref{id} and \eqref{hzjwsw} are applied.
\end{enumerate}
For further evaluations of \eqref{rs2}--\eqref{rl2}, we scrutinize
a lower bound {on} $h(Z^n|J(W),S(W))$ and {an} upper bound {on} $h(Y^n_W|J(W),S(W))$ {with} fixed $h(X^n_W|J(W),S(W))$ by applying the conditional EPI \cite[Lemma II]{bergmans1974}.
It is a key to set
\begin{align}
    \hspace{-3mm} \frac{1}{n}h(X^n_W|J(W),S(W)) = \frac{1}{2}\log\left(2\pi e (\alpha\rho^2_1 + 1 - \rho^2_1)\right) \label{xjsalpah}
\end{align}
{with some} $0 < \alpha \le 1$. Indeed, this is reasonable setting because
$\frac{1}{2}\log(2\pi e)\ge \frac{1}{n}h(X^n_W|J(W),S(W)) \ge\frac{1}{2}\log (2\pi e(1-\rho^2_1))$. The lower bound is obtained {from} $\frac{1}{n}h(X^n_W|J(W),S(W)) \ge \frac{1}{n}h(X^n_W|Y^n_W,J(W),S(W)) = \frac{1}{n}h(X^n_W|Y^n_W)$ due to the fact that $(J(W),S(W))$ {is a} function of $Y^n_W$.

{In} the direction {from} $X$ to $Z$, by applying the conditional EPI \cite[Lemma II]{bergmans1974} to the first equality in \eqref{zxni2}, it follows that
\begin{align}
    &e^{\frac{2}{n}h(Z^n|J(W),S(W))} \nonumber \\
    &~~~~~~~~~~\ge e^{\frac{2}{n}h(\rho_2X^n|J(W),S(W))} + e^{\frac{2}{n}h(N^n_2|J(W),S(W))}, \nonumber \\
    &~~~~~~~~~~\overset{\mathrm{(l)}}{=} \rho^2_2e^{\frac{2}{n}h(X^n |J(W),S(W))} + e^{\frac{2}{n}h(N^n_2)}, \nonumber \\
    &~~~~~~~~~~= \rho^2_2\left(2\pi e(\alpha\rho^2_1 + 1 - \rho^2_1)\right) + 2\pi e( 1 - \rho^2_2), \nonumber \\
    &~~~~~~~~~~= 2\pi e(\alpha\rho^2_1\rho^2_2 + 1 - \rho^2_1\rho^2_2),
\end{align}
where (l) holds as $N^n_2$ is independent of $(J(W),S(W))$, and as a deduction,
\begin{align}
    \hspace{-4mm} \frac{1}{n}h(Z^n|J(W),S(W)) \ge \frac{1}{2}\log(2\pi e(\alpha\rho^2_1\rho^2_2 + 1 - \rho^2_1\rho^2_2)). \label{zjsalpah}
\end{align}
In the opposite direction (from $X$ to $Y$), by again applying the conditional EPI \cite[Lemma II]{bergmans1974} to \eqref{xyne}, we have that
\begin{align}
    e^{\frac{2}{n}h(X^n |J(W),S(W))} &\ge e^{\frac{2}{n}h(\rho_1Y^n |J(W),S(W))} + e^{\frac{2}{n}h(N'^n_1)},
\end{align}
{meaning that}
\begin{align}
    2 \pi e (\alpha\rho^2_1 + 1 - \rho^2_1) &\ge \rho^2_1e^{\frac{2}{n}h(Y^n |J(W),S(W))} + 2 \pi e (1-\rho^2_1)
\end{align}
and thus
\begin{align}
    e^{\frac{2}{n}h(Y^n |J(W),S(W))} &\le 2 \pi e \alpha.
\end{align}
Hence, it follows that
\begin{align}
    \frac{1}{n}h(Y^n |J(W),S(W)) \le \frac{1}{2}\log(2 \pi e \alpha), \label{yjsalpah}
\end{align}
which is not derivable from the first equation in \eqref{zxne} of the original system. Now {plugging} \eqref{xjsalpah}, \eqref{zjsalpah}, and \eqref{yjsalpah} into \eqref{rs2}--\eqref{rl2}, we obtain that
\begin{align}
& R_I + R_S \leq \frac{1}{2}\log\left(\frac{1}{\alpha\rho^2_1\rho^2_2 + 1 - \rho^2_1\rho^2_2}\right) + 3\delta + \delta_n,\label{rirsfinal} \\
&R_J \geq \frac{1}{2}\log\left(\frac{\alpha\rho^2_1\rho^2_2 + 1 - \rho^2_1\rho^2_2}{\alpha}\right) + R_I - ({3\delta}+\delta_n), \label{rjfinal} \\
&R_L \ge \frac{1}{2}\log\left(\frac{\alpha\rho^2_1\rho^2_2 + 1 - \rho^2_1\rho^2_2}{\alpha\rho^2_1 + 1 - \rho^2_1}\right) + R_I - ({3\delta} + \delta_n). \label{rlfinal}
\end{align}

Eventually, by letting $n \rightarrow \infty$ and $\delta \downarrow 0$, from \eqref{rirsfinal}--\eqref{rlfinal}, we can see that the capacity region is contained in the {right-hand} side of (\ref{theorem1}).
\qed

\subsection{Achievability Part}
%There are some similarities to the achievability proof in \cite{willems2009}.
\noindent{\em Overviews}:

%We generate auxiliary sequences based on the relations of \eqref{xyne} and \eqref{zxni2} from edge $Y$.
The modified {typical} set (cf.\ Definition \ref{modifyset}), giving the so-called Markov lemma for weak typicality, and Gaussian typicality \cite[Section 8.2]{cover} help us show {that} the error probability of the BIS vanishes for large enough $n$. Though a more general version of {the} Markov lemma for Gaussian {sources}, including lossy reconstruction, is shown in \cite{oohama1997}, we found out that the two properties of the modified {typical} set are handy tools for checking all conditions in Definition \ref{def11}{,} and thus we provide our proof of the achievability based on this set. For evaluating the uniformity of secret keys \eqref{secretk}, secrecy-leakage \eqref{secrecy}, and privacy-leakage \eqref{privacy}, we extend \cite[Lemma 4]{kitti} to include continuous RVs so that the extended one can be used to derive the upper bounds on conditional differential entropies of jointly typical sequences, appearing in these evaluations.

Let $0<\alpha \le 1$ and fix $\delta>0$ (small enough positive), the block length $n$, and the joint pdf of $(U,Y,X,Z)$ such that the Markov chain $U-Y-X-Z$ holds, {where we let} $U$ be Gaussian with mean zero and variance $1-\alpha$. Now consider that
\begin{align}
    Y = U + \Phi, \label{yuphi}
\end{align}
where $\Phi$, independent of $U$, is Gaussian with mean zero and variance $\alpha$. From \eqref{xyne} and \eqref{zxni2} of the converted system, it yields that
\begin{align}
    X &= \rho_1U + \rho_1\Phi + N'_1, \label{xuphi} \\
    Z &= \rho_1\rho_2U + \rho_1\rho_2\Phi + \rho_2N'_1 + N_2. \label{zuphi}
\end{align}
Hence, we readily see that
\begin{align}
    I(Y;U) &= \frac{1}{2}\log \frac{1}{\alpha},~
    I(X;U) = \frac{1}{2}\log \left(\frac{1}{\alpha\rho^2_1 + 1 - \rho^2_1}\right), \nonumber \\
    I(Z;U) &=\frac{1}{2}\log \left(\frac{1}{\alpha\rho^2_1\rho^2_2 + 1 - \rho^2_1\rho^2_2}\right). \label{iyuxuzu}
\end{align}

Now set $0 < R_I < I(Z;U)$, and 
\begin{align}
    R_S &= I(Z;U) - R_I - 2\delta, \\
    R_J &= I(Y;U) - I(Z;U) + R_I + 6\delta, \\
    R_L &= I(X;U) - I(Z;U) + R_I + 6\delta, \\
    M_I & = 2^{nR_I},~~M_S = 2^{nR_S},~~M_J = 2^{nR_J},
\end{align}
where the values of $I(Y;U), I(X;U)$, and $I(Z;U)$ are specified in \eqref{iyuxuzu}. Also, remind that $\mathcal{I} = [1:M_I],\mathcal{S} = [1:M_S],~\mathcal{J} = [1:M_J]$.

Next we generate $2^{n(I(Y;U)+\delta)}$ sequences of $u^n(s,j)$, where each symbol of these sequences is i.i.d. Gaussian with mean zero and variance $1-\alpha$, and $s \in \mathcal{S}$ and $j \in \mathcal{J}$.

Seeing $y^n_i~(i \in \mathcal{I})$, the encoder finds $u^n(s,j)$ such that $(y^n_i,u^n({s},j)) \in {\mathcal{B}^{(n)}_{\delta}}(YU)$. {If} there are multiple pairs of such $(s,j)$, the encoder picks one at random. Otherwise, it declares error. We denote the chosen pair as $(s(i),j(i))$, where they are function of the index $i$. {Template} $j(i)$ is stored in the public DB and {secret key} $s(i)$ is saved in the key DB.

Observing $z^n$, the noisy sequence of the identified user $x^n_w$, the decoder looks for $u^n(s,j(i))$ such that $(z^n,u^n(s,j(i))) \in {\mathcal{A}^{(n)}_{\delta}}(ZU)$ for {some} $i \in \mathcal{I}$ and $s \in \mathcal{S}$. If a unique pair $(i,s)$ {is} found, it outputs $(\widehat{w},\widehat{s(w)})=(i,s)$, or else it declares error. {Finally, it compares $\widehat{s(w)}$ with $s(\widehat{w})$ in the key DB, and the authentication is successful if they match.}

Let $(J(i),S(i))$ denote the index pair chosen at the encoder based on $Y^n_i$, i.e., $(Y^n_i,U^n(S(i),J(i))) \in {\mathcal{B}^{(n)}_{\delta}}(YU)$. Furthermore, we denote $U^n(S(i),J(i))$ as $U^n_i$ for {simplicity}. Next, we check all conditions in Definition \ref{def11} {hold} for a random codebook $\mathcal{C}_n = \{U^n(s,j), s \in \mathcal{S}$ and $j \in \mathcal{J}\}$.

\medskip
\noindent{\em Analysis of Error Probability}:
For $W = i$, {an} error event possibly happens at the encoder is:
\begin{enumerate}[label={}]
	\item [$\mathcal{E}_1$]:\{$(Y^n_i,U^n(s,j)) \notin {\mathcal{B}^{(n)}_{\delta}}(YU)$  for all $s \in \mathcal{S}$ and $j \in \mathcal{J}$\},
\end{enumerate}
and {those} at the decoder are:
\begin{enumerate}[label={}]
	\item [${\mathcal{E}_2}$]:~\{$(Z^n,U^n(J(i),S(i))) \notin {\mathcal{A}^{(n)}_{\delta}}(ZU)$\},
    \item [${\mathcal{E}_3}$]:~\{$(Z^n,U^n(J(i),s'){)} \in {\mathcal{A}^{(n)}_{\delta}}(ZU)$ for some $\exists s' \neq S(i)~(s' \in \mathcal{S})$\}.
    \item [${\mathcal{E}_4}$]:~\{$(Z^n,U^n(J(i'),s'){)} \in {\mathcal{A}^{(n)}_{\delta}}(ZU)$ for some $\exists i' \neq i~(i' \in \mathcal{I})$ and $s' \in \mathcal{S}$\}.
\end{enumerate}
Note that the authentication process is guaranteed to be successful if the genuine index and secret key of the identified user are {correctly} estimated at the decoder, indicating that it is sufficient to focus on assessing the probability of incorrect {estimation} for the pair at the decoder. Then, the error probability can be further evaluated as
\begin{align}
\Pr&\{(\widehat{W},\widehat{S(W)})\neq (W,S(W))|W=i\} \nonumber \\
&= \Pr\{\mathcal{E}_1\cup\mathcal{E}_2\cup\mathcal{E}_3\cup\mathcal{E}_4\} \nonumber \\
&\leq \Pr\left\{\mathcal{E}_1\right\} + \Pr\left\{{\mathcal{E}_2|\mathcal{E}^c_1}\right\}+{\Pr\left\{\mathcal{E}_3\cup\mathcal{E}_4\right\}}. \label{errorpro}
\end{align}
By applying the similar arguments of \cite[Appendix A-B]{itw}, it can be shown that the entire error probability vanishes. Nonetheless, we provide the details for completeness of the proof.
%Here, instead of providing the details, we give hints of how each term in \eqref{errorpro} can be verified. The first and second terms can be made arbitrarily small by in a similar arguments of First and Third terms in \cite[Appendix A-B, Error Probability]{itw}, respectively. The third and forth terms in \eqref{errorpro} go to zero for large enough $n$ by the similar argument of Forth term in \cite[Appendix A-B, Error Probability]{itw}.

The first term $\Pr\left\{\mathcal{E}_1\right\}$ can be evaluated as
\begin{align*}
    \Pr&\left\{\mathcal{E}_1\right\} \nonumber \\
    &= \Pr\left[\bigcap_{s \in \mathcal{S}, j \in \mathcal{J}}(Y^n_i,U^n(s,j)) \notin {\mathcal{B}^{(n)}_{\delta}}(YU)\right] \nonumber \\
    &=\prod_{s=1}^{|\mathcal{S}|}\prod_{j=1}^{|\mathcal{J}|}\Pr\{(Y^n_i,U^n(s,j)) \notin {\mathcal{B}^{(n)}_{\delta}}(YU)\} \nonumber \\
    &\overset{\mathrm{(a)}}{=}\int f_{{Y^n_i}}(y^n)\prod_{s=1}^{|\mathcal{S}|}\prod_{j=1}^{|\mathcal{J}|}\Pr\{U^n(s,j) \notin{\mathcal{B}^{(n)}_{\delta}}(U|y^n)\}dy^n \nonumber \\
    &= \int f_{Y^n_i}(y^n)\left\{\int_{ {\mathcal{B}^{(n)}_{\delta}}(U|y^n)^c}f_{U^n}(u^n)du^n\right\}^{|\mathcal{S}\times\mathcal{J}|}dy^n \nonumber \\
    &= \int f_{{Y^n_i}}(y^n)\left(1-\int_{{\mathcal{B}^{(n)}_{\delta}}(U|y^n)}f_{U^n}(u^n)du^n\right)^{|\mathcal{S}\times\mathcal{J}|}dy^n 
\end{align*}
\begin{align}
    &\overset{\mathrm{(b)}}{\le} \int f_{{Y^n_i}}(y^n)\Bigg{(}1-2^{-n(I(U;Y) + 3\delta)}\nonumber \\
    &~~~~\cdot\int_{{\mathcal{B}^{(n)}_{\delta}}(U|y^n)}f_{U^n|{Y^n_i}}(u^n|y^n)du^n\Bigg{)}^{|\mathcal{S}\times\mathcal{J}|}dy^n \nonumber \\
    &\overset{\mathrm{(c)}}{\le} \int f_{{Y^n_i}}(y^n)\Bigg(1-\int_{{\mathcal{B}^{(n)}_{\delta}}(U|y^n)}f_{U^n|{Y^n_i}}(u^n|y^n)du^n \nonumber \\
    &~~~~+ 2^{-|\mathcal{S}\times\mathcal{J}|\cdot2^{-n(I(U;Y) + 3\delta)}}\Bigg)dy^n\nonumber \\
    &\overset{\mathrm{(d)}}{=} \iint_{ {\mathcal{B}^{(n)}_{\delta}}(U|y^n)^c}f_{U^n{Y^n_i}}(u^n,y^n)du^ndy^n \nonumber \\
    &~~~~+ 2^{-2^{n\delta}}\int f_{{Y^n_i}}(y^n)dy^n \nonumber\\
    &\overset{\mathrm{(e)}}{\le} 2\delta \label{pre1}
\end{align}
for large enough $n$, where

\begin{enumerate}[label=(\alph*)]
	\setcounter{enumi}{0}
	\item {is due to the fact that $Y^n_i$ and $U^n(s,j)$ are mutually independent,}
    \item {is obtained} by applying Property 1 of the modified {$\delta$-typical} set \cite{itw}, suggesting that if $(y^n,u^n) \in \mathcal{B}^{(n)}_{\epsilon}(YU)$, $(y^n,u^n)$ is also a member of $\mathcal{A}^{(n)}_{\epsilon}(YU)$, and thus
    \begin{align}
        f_{U^n}(u^n) &= f_{U^n|{Y^n_i}}(u^n|y^n)\frac{f_{U^n}(u^n)\cdot f_{{Y^n_i}}(y^n)}{f_{U^n{Y^n_i}}(u^n,y^n)} \nonumber \\
        & \ge f_{U^n|{Y^n_i}}(u^n|y^n)\frac{2^{-n(h(U)+\delta)}\cdot2^{-n(h(Y)+\delta)}}{2^{-n(h(Y,U)+\delta)}} \nonumber \\
        &=f_{U^n|{Y^n_i}}(u^n|y^n)2^{-n(I(Y;U) + 3\delta)},
    \end{align}
    \item follows because $(1-\alpha\beta)^m \le 1 - \alpha + 2^{-m\beta}$ \cite{cover} is applied,
    \item holds since $\frac{1}{2}\log|\mathcal{S}|+\frac{1}{2}\log|\mathcal{J}| = I(Y;U) + 4\delta$.
    \item follows by applying Property 2 of the modified {$\epsilon$-typical} set \cite{itw}.
\end{enumerate}

For the second term, it follows that
\begin{align}
    &\Pr\left\{{\mathcal{E}_2|\mathcal{E}^c_1}\right\} \nonumber \\
    &= \Pr\{(Z^n,U^n_i) \notin {\mathcal{A}^{(n)}_{\delta}}(ZU)|(Y^n_i,U^n_i) \in {\mathcal{B}^{(n)}_{\delta}}(YU)\} \nonumber \\
    &\le \Pr\{(Z^n,Y^n_i,U^n_i) \notin {\mathcal{A}^{(n)}_{\delta}}(ZYU)|(Y^n_i,U^n_i) \in {\mathcal{B}^{(n)}_{\delta}}(YU)\} \nonumber \\
    &= \iint_{{\mathcal{B}^{(n)}_{\delta}}(YU)} f_{Y^n
    _iU^n_i}(y^n,u^n) \nonumber \\
    &~~\cdot\Pr\{Z^n \notin {\mathcal{A}^{(n)}_{\delta}}(Z|y^n,u^n)|(Y^n_i,U^n_i) = (y^n,u^n)\}d(y^n,u^n) \nonumber \\
    &\overset{\mathrm{(f)}}{\le} \delta \iint_{{\mathcal{B}^{(n)}_{\delta}}(YU)} f_{Y^n
    _iU^n_i}(y^n,u^n)d(y^n,u^n) \nonumber \\
    &\le \delta, \label{pre2e1c}
\end{align}
where (f) follows from the definition of the modified {$\delta$-typical} set due to the Markov chain $Z-Y-U$.

Finally, the last term $\Pr\left\{\mathcal{E}_3\cup\mathcal{E}_4\right\}$ can be bounded as
\begin{align}
    \Pr&\left\{\mathcal{E}_3\cup\mathcal{E}_4\right\} \nonumber \\
    &= \Pr\left[\bigcup_{\substack{i' \in \mathcal{I}, s' \in \mathcal{S}}}(Z^n,U^n(s',J(i'))) \in {\mathcal{A}^{(n)}_{\delta}}(ZU)\right] \nonumber \\
    &\le \sum_{i'=1}^{|\mathcal{I}|}\sum_{s'=1}^{|\mathcal{S}|}\Pr\left\{(Z^n,U^n(s',J(i'))) \in {\mathcal{A}^{(n)}_{\delta}}(ZU)\right\} \nonumber \\
    &\le \sum_{i'=1}^{|\mathcal{I}|}\sum_{s'=1}^{|\mathcal{S}|} 2^{-n(I(Z;U)-\delta)} \nonumber \\
    &= |\mathcal{I}\times\mathcal{S}|\cdot 2^{-n(I(Z;U)-\delta)} \nonumber \\
    &\overset{\mathrm{(g)}}{=} 2^{-n\delta}, \label{pre3e4}
\end{align}
where (g) follows as $\frac{1}{n}\log M_I + \frac{1}{n}\log M_S = I(Z;U)-2\delta$.

Consequently,
\begin{align}
    \Pr&\{(\widehat{W},\widehat{S(W)})\neq (W,S(W))|W=i\} \le {4\delta} \label{errorwsw}
\end{align}
for large enough $n$.

Before proceeding further, we introduce a lemma that is often used in the sequel. Again recall that the index pair $(J(i),S(i))$ determines the chosen sequence $U^n_i$ directly and thus the following lemma can be thought of an extended version of \cite[Lemma 4]{kitti}, incorporating continuous RVs.
\begin{Lemma}
It holds that
\begin{align}
    \frac{1}{n}h(Y^n_i|S(i),J(i),\mathcal{C}_n) &\le h(Y|U) + \delta_n, \label{1122} \\
    \frac{1}{n}h(Y^n_i|X^n_i,J(i),S(i),\mathcal{C}_n) &\le h(Y|X,U) + \delta_n, \label{1123}
\end{align}
where $\delta_n \downarrow 0$ as $\delta \downarrow 0$ and $n \rightarrow \infty$.
\end{Lemma}

\noindent{Proof}:~~~~The tie between the modified {$\delta$-typical} set ${\mathcal{B}^{(n)}_{\delta}}(\cdot)$ and the weakly $\delta$-typical set ${\mathcal{A}^{(n)}_{\delta}}(\cdot)$ is helpful for proving the above lemma.
We first prove \eqref{1122}.

Define {an} RV $T$ as follows:
\begin{align}
T =
  \begin{cases}
    1~~{ \rm if}~~(Y^n_i,U^n_i) \in {\mathcal{B}^{(n)}_{\delta}}(YU), \\
    0~~{ \rm otherwise}.
  \end{cases}
\end{align}

In the analysis of {the} error probability, we have already demonstrated that $P_{T}(0) \le 2\delta$, or $(Y^n_i,U^n_i) \in {\mathcal{B}^{(n)}_{\delta}}(YU)$ with high probability. From the left-hand side of \eqref{1122},
\begin{align}
    h(Y^n_i|J(i),&S(i),\mathcal{C}_n) \nonumber \\
    &\overset{\mathrm{(h)}}{{=}} h(Y^n_i|U^n_i,J(i),S(i),\mathcal{C}_n) \nonumber \\
    &\overset{\mathrm{(i)}}{\le} h(Y^n_i|U^n_i)\le h(Y^n_i,T|U^n_i) \nonumber \\
    &\le H(T) + h(Y^n_i|U^n_i,T) \nonumber \\
    &= 1 + P_T(0)h(Y^n_i|U^n_i,T=0) \nonumber \\
    &~~~+ P_T(1)h(Y^n_i|U^n_i,T=1) \nonumber \\
    &\overset{\mathrm{(j)}}{\le} n\epsilon_n + h(Y^n_i|U^n_i,T=1) \nonumber \\
    &= n\epsilon_n +  \iint_{{\mathcal{B}^{(n)}_{\delta}}(YU)}f_{Y^n_iU^n_i}(y^n,u^n)\nonumber \\
    &~~~\cdot \log \frac{\Pr\{(Y^n_i,U^n_i) \in {\mathcal{B}^{(n)}_{\delta}}(YU)\}}{f_{Y^n_i|U^n_i}(y^n|u^n)}d(y^n,u^n) \nonumber \\
    %&\le n\epsilon_n +  \frac{1}{1-2\delta}\int_{u^n_i}\int_{y^n_i \in {\mathcal{B}^{(n)}_{\delta}}(Y|u^n_i)}f_{Y^n_iU^n_i}(y^n_i,u^n_i) \nonumber \\
    %&~~~\cdot \log \frac{1}{f_{Y^n_i|U^n_i}(y^n_i|u^n_i)}dy^n_idu^n_i \nonumber \\
    &\overset{\mathrm{(k)}}{\le} n\epsilon_n + n(h(Y|U) + 2\delta)) \nonumber \\
    &~~~\cdot\iint_{{\mathcal{B}^{(n)}_{\delta}}(YU)}f_{Y^n_iU^n_i}(y^n,u^n)d(y^n,u^n) \nonumber \\
    &\le n(h(Y|U) + 2\delta + \epsilon_n), \label{yjscn}
\end{align}
where
\begin{enumerate}[label=(\alph*)]
	\setcounter{enumi}{7}
        \item follows as $(J(i),S(i))$ determines $U^n_i$,
        \item follows because conditioning reduces entropy,
        \item  {follows as $h(Y^n_i|U^n_i,T=0) \le h(Y^n_i) = \frac{n}{2}\log(2 \pi e)$, and we define $\epsilon_n = \frac{1}{n} + \delta \log(2 \pi e)$},
        %and $1-2\delta \le \Pr\{(Y^n_i,U^n_i) \in {\mathcal{B}^{(n)}_{\delta}}(YU)\} \le 1$ for large enough $n$,
        \item follows since {$\Pr\{(Y^n_i,U^n_i) \in {\mathcal{B}^{(n)}_{\delta}}(YU)\} \le 1$,  and from Property 1 of the modified {$\delta$-typical} set \cite{itw}, we have that} $f_{Y^n_i|U^n_i}(y^n|u^n)=\frac{f_{Y^n_iU^n_i}(y^n,u^n)}{f_{U^n_i}(u^n)} \ge \frac{2^{-n(h(Y,U)+\delta)}}{2^{-n(h(U)-\delta)}}
        = 2^{-n(h(Y|U)+2\delta)}.$
\end{enumerate}
Therefore, from \eqref{yjscn}, we obtain that
\begin{align}
    \frac{1}{n}h(Y^n_i|J(i),S(i),\mathcal{C}_n) &\le h(Y|U) + \delta_n,
\end{align}
{where $\delta_n = 2\delta + \epsilon_n$ and $\delta_n \downarrow 0$ as $n \rightarrow \infty$ and $\delta\downarrow 0$}.

Next, we briefly summarize how to show \eqref{1123}. The left-hand side of \eqref{1123} can be developed as $h(Y^n_i|X^n_i,J(i),S(i),\mathcal{C}_n) = h(Y^n_i|X^n_i,U^n_i,J(i),S(i),\mathcal{C}_n)\le h(Y^n_i|X^n_i,U^n_i,\mathcal{C}_n)$, where the first {equality} and second inequality follow due to the same reasons of {(h) and (i)} in \eqref{yjscn}, respectively. By the definition of the modified {$\delta$-typical} set \cite{itw}, it can be concluded that $\Pr\{(X^n_i,Y^n_i,U^n_i) \in {\mathcal{A}^{(n)}_{\delta}}(XYU)\} \rightarrow 1$ as $n \rightarrow \infty$ {(cf.\ \eqref{pre2e1c})} due to {the Markov chain $X-Y-U$ and} $(Y^n_i,U^n_i) \in {\mathcal{B}^{(n)}_{\delta}}({Y}U)$ {with high probability}. This implies $\Pr\{(X^n_i,U^n_i) \in {\mathcal{A}^{(n)}_{\delta}}(XU)\} \rightarrow 1$ and $\Pr\{Y^n_i \in \mathcal{A}^{(n)}_{\delta}(Y|{x^n},{u^n})|(X^n_i,U^n_i)=({x^n},{u^n})\} \rightarrow 1$ as $n \rightarrow \infty$ as well. Based on this observation, the rest of proof for \eqref{1123} can be done similarly by the arguments seen in \cite[Appendix C]{kitti}, {and} therefore the details are omitted.

\medskip
\noindent{\em Analysis of Identification and Storage Rates}:

{Equations} (\ref{id}) and \eqref{storage} obviously hold from the parameter settings.

\medskip
\noindent{\em Analysis of Secrecy Rate}:
\begin{align}
    H(S(i)|\mathcal{C}_n) &= h(Y^n_i,J(i),S(i)|\mathcal{C}_n) - H(J(i)|S(i),\mathcal{C}_n) \nonumber \\
    &~~~~- h(Y^n_i|J(i),S(i),\mathcal{C}_n) \nonumber \\
    &\overset{\mathrm{(l)}}{\ge} h(Y^n_i) - H(J(i)|\mathcal{C}_n) \nonumber \\
    &~~~~- h(Y^n_i|J(i),S(i),\mathcal{C}_n) \nonumber \\
    &\overset{\mathrm{(m)}}{\ge} nh(Y) - n(I(Y;U)-I(Z;U)+R_I+6\delta) \nonumber \\
    &~~~~- n(h(Y|U) + \delta_n) \nonumber \\
    &= n(I(Z;U) - R_I - 6\delta - \delta_n) \nonumber \\
    &= n(R_S - 4\delta - \delta_n).
\end{align}
where
\begin{enumerate}[label=(\alph*)]
	\setcounter{enumi}{11}
        \item {follows because $(J(i),S(i))$ is a function of $Y^n_i$},
        \item follows as $\frac{1}{n}H(J(i)|\mathcal{C}_n) \le I(Y;U)-I(Z;U)+R_I+{6\delta}$ and \eqref{1123} is applied.
\end{enumerate}
Thus,
\begin{align}
    \frac{1}{n}H(S(i)|\mathcal{C}_n) \ge R_S - 5\delta = \frac{1}{n}\log M_S - 5\delta \label{hsi}
\end{align}
for large enough $n$.

\medskip
\noindent{\em Analysis of Secrecy-Leakage}:
\begin{align}
    &I(S(i);J(i)|\mathcal{C}_n) \nonumber \\
    &= H(S(i)|\mathcal{C}_n) + H(J(i)|\mathcal{C}_n) - h(Y^n_i,S(i),J(i)|\mathcal{C}_n) \nonumber \\
    &~~~+ h(Y^n_i|S(i),J(i),\mathcal{C}_n) \nonumber \\
    &\overset{\mathrm{(n)}}{\le} n(I(Z;U)-R_I-2\delta + I(Y;U)-I(Z;U) + R_I + 6\delta) \nonumber \\
    &~~~- nh(Y)+ n(h(Y|U) + \delta_n) \nonumber \\
    &=n(4\delta + \delta_n),
\end{align}
where (n) follows because $\frac{1}{n}H(S(i)|\mathcal{C}_n) \le I(Z;U)-R_I-{2\delta}$, $\frac{1}{n}H(J(i)|\mathcal{C}_n) \le I(Y;U)-I(Z;U)+R_I+{6\delta}$, and \eqref{1123} is applied.
Hence,
\begin{align}
    \frac{1}{n}I(S(i);J(i)|\mathcal{C}_n) \le 5\delta \label{isiji}
\end{align}
for sufficiently large $n$.

\medskip
\noindent{\em Analysis of Privacy-Leakage Rate}: From the left-hand side of \eqref{privacy}, we have that
\begin{align}
    I&(X^n_i;J(i)|\mathcal{C}_n) \nonumber\\
    &= H(J(i)|\mathcal{C}_n) - H(J(i)|X^n_i,\mathcal{C}_n) \nonumber\\
    &\le n(I(Y;U)-I(Z;U)+R_I + 6\delta) - H(J(i)|X^n_i,\mathcal{C}_n) \nonumber\\
    &= n(h(U|Z)-h(U|Y)+R_I + 6\delta) - H(J(i)|X^n_i,\mathcal{C}_n). \label{ixnj}
\end{align}
The last term in \eqref{ixnj} can be further evaluated as
\begin{align}
    H(J(i)&|X^n_i,\mathcal{C}_n)  \nonumber\\
    &= h(Y^n_i,J(i)|X^n_i,\mathcal{C}_n) - h(Y^n_i|X^n_i,J(i),\mathcal{C}_n) \nonumber\\
    &= h(Y^n_i|X^n_i,\mathcal{C}_n) - h(Y^n_i|X^n_i,J(i),S(i),\mathcal{C}_n)  \nonumber\\
    &~~~- H(S(i)|J(i),X^n_i,\mathcal{C}_n) \nonumber\\
    &\overset{\mathrm{(o)}}{=} nh(Y|X) - h(Y^n_i|X^n_i,J(i),S(i),\mathcal{C}_n) \nonumber \\
    &~~~- H(S(i)|J(i),X^n_i,Z^n,\mathcal{C}_n) \nonumber \\
    &\overset{\mathrm{(p)}}{\ge} nh(Y|X) - h(Y^n_i|X^n_i,J(i),S(i),\mathcal{C}_n)  \nonumber\\
    &~~~- H(S(i)|\bm{J},Z^n,\mathcal{C}_n) \nonumber\\
    &\overset{\mathrm{(q)}}{\ge} nh(Y|X) - h(Y^n_i|X^n_i,J(i),S(i),\mathcal{C}_n) - n\delta''_n \nonumber\\
    &\overset{\mathrm{(r)}}{\ge} nh(Y|X) - n(h(Y|X,U) + \delta_n) - n\delta''_n \nonumber\\
    &= n(I(Y;U|X)-\delta_n-\delta''_n)  \nonumber\\
    &= n(h(U|X) - h(U|Y)-\delta_n-\delta''_n) \label{ixnj22}
\end{align}
where
\begin{enumerate}[label=(\alph*)]
	\setcounter{enumi}{14}
        \item follows since $Y^n_i$ and $X^n_i$ are independent of $\mathcal{C}_n$ and the Markov chain $S(i)-(J(i),X^n_i)-Z^n$ holds,
        \item follows because conditioning reduces entropy and $S(i)-(J(i),Z^n)-\bm{J}\backslash J(i)$ is applied,
        \item follows by applying Fano's inequality, and $\delta''_n\downarrow 0$ as $\delta\downarrow 0$ and $n \rightarrow \infty$,
        \item {is} due to \eqref{1123}.
\end{enumerate}
From \eqref{ixnj} and \eqref{ixnj22}, we have that
\begin{align}
    \hspace{-3mm}\frac{1}{n}I(X^n_i;J(i)|\mathcal{C}_n) &\le h(U|Z) - h(U|X) + R_I + 6\delta + \delta_n + \delta''_n \nonumber\\
    &= I(X;U) - I(Z;U) + R_I + 6\delta + \delta_n + \delta''_n \nonumber\\
    &\le R_L + \delta \label{ixnj33}
\end{align}
for sufficiently large $n$.

Finally, by using the selection lemma \cite[Lemma 2.2]{BB}, there exists
at least a good codebook satisfying all the conditions in Definition \ref{def11} for large enough $n$.
\qed

\section{Proof Sketch of the Region $\mathcal{R}_C$}

In this section, we highlight the proof of the chosen-secret BIS model. Some parts follow from the arguments {in} Section IV, so we omit the similarities.

\subsection{Converse Part}
As seen in the converse proof of the generated-secret BIS model, we also consider the case in which $W$ is uniformly distributed on $\mathcal{I}$. Suppose that {a} pair $(R_I,R_S,R_J,R_L)$ is achievable.

For the analyses of identification, secrecy, and privacy-leakage rates,
the reader should refer to the discussions around \eqref{rs2} and \eqref{rl2}. We argue only the bound of $R_J$, which is different from the one seen in the generated-secret BIS model.

\medskip
\noindent{\em Analysis of Storage Rate}:
\begin{align}
    n(R_J + \delta) &\ge \log M_J \ge \max_{w \in \mathcal{I}}H(w) \ge H(J(W)|W) \nonumber \\
    &\overset{\mathrm{(a)}}{=}I(Y^n_W,S(W);J(W)|W) \nonumber \\
    &= h(Y^n_W,S(W)) - h(Y^n_W,S(W)|J(W)) \nonumber \\
    &\overset{\mathrm{(b)}}{=} h(Y^n_W) + H(S(W)) - H(S(W)|J(W))  \nonumber \\
    &~~~- h(Y^n_W|J(W),S(W)) \nonumber \\
    &\overset{\mathrm{(c)}}{\ge} \frac{n}{2}\log(2 \pi e) - \frac{n}{2}\log(2 \pi e \alpha) \nonumber \\
    &\ge \frac{n}{2}\log\left(\frac{1}{\alpha}\right),
\end{align}
where
\begin{enumerate}[label=(\alph*)]
	\setcounter{enumi}{0}
        \item follows since $J(W)$ is a function of $(Y^n_W,S(W))$,
        \item follows as $S(W)$ is chosen independently of $Y^n_W$,
        \item follows because conditioning reduces entropy and \eqref{yjsalpah} is applied. 
\end{enumerate}
Then, we have that
\begin{align}
    R_J \ge \frac{1}{2}\log\left(\frac{1}{\alpha}\right)-\delta.
\end{align}

By letting $n \rightarrow \infty$ and $\delta \downarrow 0$, the capacity region of the chosen-secret BIS model is contained in the right-hand side of (\ref{theorem2}).
\qed

\subsection{Achievability Part}

In order to avoid {confusion} in the subsequent arguments, we define some new notations used only in this part. The pairs $(J_C(i),S_C(i))$ and $(J_G(i),S_G(i))$ denote the helper data and secret key of individual $i$ for chosen- and generated-secret BIS models, respectively. Moreover, $M_{J_C}$ and $M_{J_G}$ denote the number of templates\footnote{Normally, $J_C(i)$, $S_C(i)$, and $M_{J_C}$ are denoted by $J(i)$, $S(i)$, and $M_J$ in other sections of this paper.}, and $R_{J_G}$ and $R_{J_C}$ denote the storage rates in the generated- and chosen-secret BIS models, respectively.

\begin{figure*}[!t]
 \begin{center}
  \includegraphics[width = 170mm]{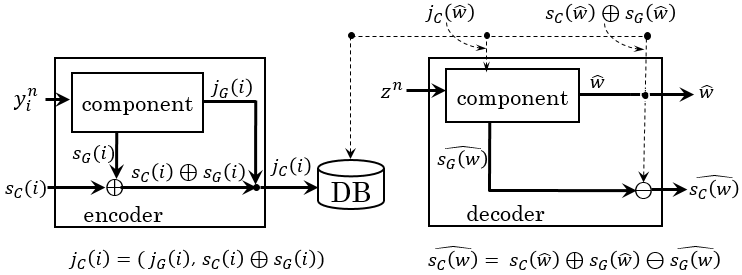}
 \end{center}
 \caption{Encoder and decoder of the chosen-secret BIS model}
 \label{fig4}
\end{figure*}

\medskip
\noindent{\em Overviews}:

The proof is an adapted version of the achievability proof of Section IV. The difference is that the encoder and decoder of the generated-secret BIS model are used as components inside the encoder and decoder of the chosen-secret BIS model as shown in Fig.\ \ref{fig4}. For encoding {$Y^n_i$ for each} user $i \in \mathcal{I}$, {the component encoder uses} a masking layer (one-time pad operation) to mask $s_C(i)$ by using $s_G(i) \in \cal{S}$ as $s_C(i)\oplus s_G(i)$, {where $\oplus$ denotes the addition modulo $M_S$}. The helper data $j_C(i)$ is the combined information of $j_G(i)$ and the masked data $s_C(i)\oplus s_G(i)$, i.e.,
\begin{align}
j_C(i) = (j_G(i),s_C(i)\oplus s_G(i)). \label{jg}
\end{align}
For decoding the identified user $w$, it first uses the component {decoder} to estimate ($\widehat{w},\widehat{s_G(w)}$) and then the secret key is retrieved by
\begin{align}
\widehat{s_C(w)} = s_C(\widehat{w})\oplus s_G(\widehat{w})\ominus \widehat{s_G(w)}, \label{sc+sc}
\end{align}
where $\ominus$ denotes the subtraction modulo $M_S$.
This technique is also used in \cite{itw}, \cite{onur}, and \cite{vy}.

Let $0<\alpha \le 1$. Fix a block length $n$ and the joint pdf of $(U,Y,X,Z)$ such that the Markov chain $U-Y-X-Z$ holds, {where} $U$ is Gaussian with mean zero and variance $1-\alpha$. The connection among the auxiliary RV $U$ and $(Y,X,Z)$ is exactly the same as the arguments around \eqref{yuphi}--\eqref{iyuxuzu}.

Now {we fix} $0 < R_I < I(Z;U)$, and 
\begin{align}
    R_S &= I(Z;U) - R_I - 2\delta, \nonumber \\
    R_{J_G} &= I(Y;U) - I(Z;U) + R_I + 6\delta, \nonumber \\
    R_{J_C} &= I(Y;U) + 3\delta, \nonumber \\
    R_L &= I(X;U) - I(Z;U) + R_I + 6\delta, \nonumber \\
    M_I & = 2^{nR_I},~~M_S = 2^{nR_S},~~M_{J_G} = 2^{nR_{J_G}},
\end{align}
where $I(Y;U), I(X;U)$, and $I(Z;U)$ are specified in \eqref{iyuxuzu}.

Next we generate $2^{n(I(Y;U)+\delta)}$ sequences of $u^n(s,j)$, where each symbol of these sequences is i.i.d. Gaussian with mean zero and variance $1-\alpha$, and $s \in \mathcal{S}$ and $j \in [1:M_{J_G}]$.

Seeing $y^n_i~(i \in \mathcal{I})$, the component {encoder looks for} $u^n(s,j)$ such that $(y^n_i,u^n(s,j)) \in {\mathcal{B}^{(n)}_{\delta}}(YU)$. {If} there are multiple pairs of such $(s,j)$, the encoder picks one at random. We denote the pair chosen by the component as $(s_G(i),j_G(i))$ and it is shared with the encoder. The encoder uses {$s_G(i)$} to conceal the chosen secret key $s_C(i)$ by $s_C(i)\oplus s_G(i)$. This masked information is combined with $j_G(i)$ to form the helper data $j_C(i)$ as $j_C(i) = \left(j_G(i),s_C(i)\oplus s_G(i)\right)$. If there does not exist such {a} pair, the component shares $(1,1)$ with the encoder. In this case, the encoder declares error.

Observing $z^n$, the noisy version of $x^n_w$, the component {decoder} looks for $u^n(s,j(i))$ such that $(z^n,u^n(s,j(i))) \in {\mathcal{A}^{(n)}_{\delta}}(ZU)$ for all $i \in \mathcal{I}$ and some $s \in \mathcal{S}$. If a unique pair $(i,s)$ {is} found, the component sets $(\widehat{w},\widehat{s_G(w)})=(i,s)$ and forwards this result to the decoder of the chosen-secret BIS model.
The decoder detects $s_C(\widehat{w})\oplus s_G(\widehat{w})$ from the public DB based on $\widehat{w}$, and outputs $\widehat{w} = i$ and $\widehat{s_C(w)}=s_C(\widehat{w})\oplus s_G(\widehat{w})\ominus \widehat{s_G(w)}$. {In the final step, the estimated key $\widehat{s_C(w)}$ and $s_C(\widehat{w})$ in the key DB are compared, and if they are equal, the authentication is successful}. If there is no such unique pair, the component shares $(1,1)$ with the decoder and upon receiving these information, error is declared.

\medskip
\noindent{{\em Analysis of Error Probability}}:

For individual $W=i$, the operation at the decoder (\ref{sc+sc}) means that $\widehat{S_C(i)} = S_C(i)$ if and only if $\widehat{S_G(i)} = S_G(i)$. In \eqref{errorwsw}, it was revealed that
$
\Pr\{(\widehat{W},\widehat{S_G(W)}) \neq (W,S_G(W))|W = i\} \leq 4\delta.
$
Therefore, the error probability of the chosen-secret BIS model can {also} be bounded by
\begin{align}
\Pr\{(\widehat{W},\widehat{S_C(W)}) \neq (W,S_C(W))|W = i\} \leq 4\delta \label{errorcsbis}
\end{align}
for large enough $n$.

\medskip
\noindent{{\em Analyses of {Identification and Secrecy Rates}}}:

{Equations} (\ref{id}) and (\ref{secretk}) are straightforward from
the parameter settings.

\medskip
\noindent{{\em Analysis of {Storage Rate}}}:
\begin{align}
    \frac{1}{n}&\log {M_{J_C}} \nonumber \\
    &\le \frac{1}{n}\log M_{J_G} + \frac{1}{n}\log M_S \nonumber \\
    &= I(Y;U) - I(Z;U) + R_I + 6\delta + I(Z;U) - R_I -2\delta \nonumber \\
    &= I(Y;U) + 4\delta = \frac{1}{2}\log \left(\frac{1}{\alpha}\right) + 4\delta \nonumber \\
    &= R_{J_C} + \delta.
\end{align}

\medskip
\noindent{{\em Analysis of Secrecy-Leakage}}:
It holds that
\begin{align}
I&(J_C(i);S_C(i)|\mathcal{C}_n) \nonumber \\
&= I(J_G(i),S_C(i)\oplus S_G(i);S_C(i)|\mathcal{C}_n) \nonumber \\
&= I(J_G(i);S_C(i)|\mathcal{C}_n) \nonumber \\
&~~~~+ I(S_C(i)\oplus S_G(i);S_C(i)|J_G(i),\mathcal{C}_n) \nonumber \\
&= I(J_G(i);S_C(i)|\mathcal{C}_n) + H(S_C(i)\oplus S_G(i)|{J_G(i),}\mathcal{C}_n) \nonumber \\
&~~~~- H(S_C(i)\oplus S_G(i)|J_G(i),S_C(i),\mathcal{C}_n) \nonumber \\
&\le I(J_G(i);S_C(i)|\mathcal{C}_n) + \log M_S \nonumber \\
&~~~~- H(S_G(i)|J_G(i),S_C(i),\mathcal{C}_n) \nonumber \\
&\overset{\mathrm{(a)}}{=} {\log M_S - H(S_G(i)|J_G(i),\mathcal{C}_n)} \nonumber \\
&= I(J_G(i);S_G(i)|\mathcal{C}_n) + \log M_S - H(S_G(i)|\mathcal{C}_n), \label{jcsc111}
\end{align}
{where (a) holds because $S_C(i)$ is chosen independently of  $(S_G(i),J_G(i))$ for given $\mathcal{C}_n$}. In \eqref{hsi} and \eqref{isiji} of Section IV, it was clarified that
\begin{align}
H(S_G(i)|\mathcal{C}_n) &\ge \log M_S - 5n\delta, \label{sc} \\
I(J_G(i);S_G(i)|\mathcal{C}_n) &\leq 5n\delta \label{jcsc}
\end{align}
for large enough $n$. Substituting (\ref{sc}) and (\ref{jcsc}) into (\ref{jcsc111}), the secrecy-leakage of the chosen-secret BIS model is bounded by
\begin{align}
\frac{1}{n}I(J_C(i);S_C(i)|\mathcal{C}_n)\leq 10\delta \label{IJSc}
\end{align}
for large enough $n$.

\medskip
\noindent{{\em Analysis of Privacy-Leakage {Rate}}}:

It can be proved that
\begin{align}
{I}(X^n_i;J_C(i)|\mathcal{C}_n)= {I}(X^n_i;J_G(i)|\mathcal{C}_n). \label{xjc}
\end{align}
{To verify this, first} one can easily see that
\begin{align}
    I(X^n_i;J_C(i)|\mathcal{C}_n)
    &= I(X^n_i;J_G(i),S_C(i)\oplus S_G(i)|\mathcal{C}_n) \nonumber \\
    %&= I(X^n_i;J_G(i)|\mathcal{C}_n) + I(X^n_i;S_C(i)\oplus S_G(i)|J_G(i),\mathcal{C}_n) \nonumber \\
    &\ge {I}(X^n_i;J_G(i)|\mathcal{C}_n). \label{xjc1}
\end{align}
{Meanwhile}, it can be shown that
\begin{align}
I&(X^n_i;J_C(i)|\mathcal{C}_n) \nonumber \\
&= I(X^n_i;J_G(i),S_C(i)\oplus S_G(i)|\mathcal{C}_n) \nonumber \\
&= I(X^n_i;J_G(i)|\mathcal{C}_n) + I(X^n_i;S_C(i)\oplus S_G(i)|J_G(i),\mathcal{C}_n) \nonumber \\
&= I(X^n_i;J_G(i)|\mathcal{C}_n) + H(S_C(i)\oplus S_G(i)|J_G(i),\mathcal{C}_n) \nonumber\\
&~~~~- H(S_C(i)\oplus S_G(i)|X^n_i,J_G(i),\mathcal{C}_n)  \nonumber \\
&\overset{\mathrm{(b)}}{\le} I(X^n_i;J_G(i)|\mathcal{C}_n) + \log M_S \nonumber\\
&~~~~- H(S_C(i)\oplus S_G(i)|X^n_i,J_G(i),S_G(i),\mathcal{C}_n)  \nonumber \\
&{=} I(X^n_i;J_G(i)|\mathcal{C}_n) + \log M_S  \nonumber \\
&~~~~- H(S_C(i)|X^n_i,J_G(i),S_G(i),\mathcal{C}_n)  \nonumber \\
&\overset{\mathrm{(c)}}{=} I(X^n_i;J_G(i)|\mathcal{C}_n) + \log M_S - \log M_S  \nonumber \\
&= {I}(X^n_i;J_G(i)|\mathcal{C}_n), \label{xjc2}
\end{align}
where
\begin{enumerate}[label=(\alph*)]
	\setcounter{enumi}{1}
        \item follows as conditioning reduces entropy,
        \item follows because $S_C(i)$ is chosen uniformly from $\mathcal{S}$ and independent of other RVs. 
\end{enumerate}
From \eqref{xjc1} and \eqref{xjc2}, \eqref{xjc} clearly holds.
By invoking the result of \eqref{ixnj33}, the privacy-leakage rate can also be made that
\begin{align}
\frac{1}{n}{I}(X^n_i;J_C(i)|\mathcal{C}_n) &\leq R_L + \delta \label{1_1}
\end{align}
for large enough $n$.

Finally, by using the selection lemma \cite[Lemma 2.2]{BB}, there is
at least {one} good codebook satisfying all the conditions in Definition \ref{def22} for large enough $n$.
\qed

\section{Conclusion and Future Work}
We characterized the capacity region of identification, secrecy, storage, and privacy-leakage rates for {both generated- and chosen-secret BIS models} under Gaussian sources. The models considered in this study are the RSM, namely, the enrollment channel is noisy. We showed that an idea for deriving the capacity regions is to convert the system to another one where the data flows of each user are in one-way direction. We also gave numerical computations of three different examples for {the derived regions} and from these results, it appeared that achieving high secrecy and small privacy-leakage rates simultaneously is unlikely manageable. For future work, we plan to extend this scenario to consider Gaussian vector sources and channels.

\end{document}